\documentclass[12pt,preprint]{aastex}
\newcommand{\beq}{\begin{equation}}
\newcommand{\beqa}{\begin{eqnarray}}
\newcommand{\eeq}{\end{equation}}
\newcommand{\eeqa}{\end{eqnarray}}
\newcommand{\simg}{\gtrsim}
\newcommand{\siml}{\lesssim}

\shorttitle{CP from GRB Afterglows}
\shortauthors{Matsumiya \& Ioka}
\begin{document}
\title{
Circular Polarization from Gamma-Ray Burst Afterglows
}
\author{
Makoto Matsumiya and Kunihito Ioka
}
\affil{Department of Earth and Space Science, Graduate School of
Science, Osaka University, Toyonaka 560-0043, Japan}
\email{matumiya@vega.ess.sci.osaka-u.ac.jp}
\email{ioka@vega.ess.sci.osaka-u.ac.jp}

\begin{abstract}
We investigate the circular polarization (CP) 
from Gamma-Ray Burst (GRB) afterglows.
We show that a tangled magnetic field cannot generate
CP without an ordered field because
there is always an oppositely directed field, so that
no handedness exists.
This implies the observation of CP
could be a useful probe of an ordered field,
which carries valuable information on the GRB central engine.
By solving the transfer equation of polarized radiation,
we find that the CP reaches 
$1\%$ at radio frequencies 
and $0.01\%$ at optical for the forward shock,
and $10$-$1\%$ at radio and $0.1$-$0.01\%$ at optical
for the reverse shock.

\end{abstract}


\section{INTRODUCTION AND SUMMARY}
Recently a very large linear polarization (LP), $\sim 80\pm 20\%$,
in the prompt $\gamma$-ray emission
of GRB 021206 was discovered (Coburn \& Boggs 2003).
The degree of LP was at the theoretical maximum 
of the synchrotron emission,
which implies an uniform ordered magnetic field 
over the visible region (but see also Eichler \& Levinson 2003 for the
scattering origin of LP).
Since the causally connected region is smaller than the visible one
(Gruzinov \& Waxman 1999),
an ordered field could be advected from the central engine
of the Gamma-Ray Burst (GRB)
and even drive the GRB explosion (Coburn \& Boggs 2003;
Lyutikov, Pariev \& Blandford 2003).

On the other hand, LP of $\siml 10\%$ (typically a few $\%$) has
been detected in the GRB afterglows
(Covino et al. 2003 and references there in),
which is attributed to synchrotron emission behind a shock
(e.g., M${\acute {\rm e}}$sz${\acute {\rm a}}$ros 2002).
In most popular models 
(Gruzinov 1999; Ghisellini \& Lazzati 1999; Sari 1999; Rossi et al. 2002),
the magnetic field is generated at the shock front 
and completely tangled (Medvedev \& Loeb 1999).
LP arises due to the geometric asymmetry
provided by the afterglow jet observed off-axis
if the magnetic fields parallel and perpendicular to the jet
have different strengths.
The $\gamma$-ray LP mentioned above could be also explained
in this model if the jet is very narrow 
(Waxman 2003; Nakar, Piran \& Waxman 2003; but see also Granot 2003).

Thus the present issue is {\it whether an ordered magnetic field exists
or not}.
If an ordered field exists in afterglows, 
its fraction to a tangled field carries 
valuable information on the GRB central engine.
In this Letter we show 
that observations of the circular polarization (CP)
could be a useful probe of the ordered field.
CP has been detected in AGN jets 
(Wardle et al. 1998; Homan \& Wardle 1999; Bower et al. 2002)
and X-ray binaries (Fender, et al. 2000, 2002) in recent years.
Theoretically, these observations are explained 
by a plasma effect in synchrotron sources. 
We apply this theory to the GRBs for the first time.

There are two main mechanisms to generate CP:
intrinsic CP of synchrotron emission
and Faraday conversion (FC) in sources.
FC is a plasma effect which converts LP into CP 
(e.g., Jones \& O'Dell 1977a,b).
These are treated all together
by solving the transfer equation
of polarized radiation in \S \ref{sec:trans}.
Then, we show that {\it the tangled field cannot generate CP}
in \S \ref{sec:mag}.
Next we estimate CP from GRB afterglows 
in presence of an ordered field together with a tangled field
in \S \ref{sec:after}.
We find that,
if the ordered field is comparable to or more than the tangled one,
the degree of CP is about $1\%$ at radio frequencies 
and $0.01\%$ at optical for the forward shock,
and it reaches $10$-$1\%$ at radio and $0.1$-$0.01\%$ at optical
for the (early) reverse shock.
The radio CP of the reverse shock remains to be $\sim 1\%$
even if the ordered field is weak ($1\%$ of the tangled one).
CP in reverse shock radio emission of GRB 990123
has an upper limit of $37\%$ at the $99.9\%$ confidence level
(Kulkarni et al. 1999; see also Finkelstein, Ipatov \& Gnedin 2002).
Further observation of CP will be a diagnosis of the ordered field
and bring clues to the nature of the GRBs.

\section{TRANSFER EQUATION}\label{sec:trans}
We first consider the evolution of the Stokes parameters, 
$I,Q,U,V$, (in units of erg s$^{-1}$ cm$^{-2}$ sr$^{-1}$ Hz$^{-1}$)
in a homogeneous plasma with a weakly anisotropic dielectric tensor 
(Sazonov \& Tsytovich 1968; Sazonov 1969a,b; Jones \& O'Dell 1977a; 
Melrose 1980).
It is described by the transfer equation of 
the polarized radiation,
\begin{eqnarray}
\left(
\begin{array}{cccc}
d/ds+\kappa_I & \kappa_Q \cos 2\phi & -\kappa_Q \sin 2\phi & \kappa_V \\
\kappa_Q \cos2\phi & d/ds+\kappa_I & \kappa^*_V & \kappa^*_Q \sin 2\phi \\
-\kappa_Q \sin 2\phi & -\kappa^*_V & d/ds+\kappa_I & \kappa^*_Q \cos 2\phi \\
\kappa_V & -\kappa^*_Q \sin 2\phi & -\kappa^*_Q \cos 2\phi & d/ds+\kappa_I \\
\end{array}     \right) 
 \left(
\begin{array}{c}
I\\
Q\\
U\\
V\\ 
\end{array}     \right)
&=&
\left(
\begin{array}{c}
\eta_I \\
\eta_Q \cos 2\phi\\
-\eta_Q \sin 2\phi\\
\eta_V \\
\end{array}     \right),
\label{eq:trans}
\end{eqnarray}
where $\phi$ is the azimuthal projection angle of the
magnetic field on the plane perpendicular to the line of sight
(see Figure~\ref{fig:magdir}),
$\kappa_{(I,Q,U,V)}$ are the absorptivity,
$\eta_{(I,Q,U,V)}$ are the emissivity, and $\kappa^*_Q$ and
$\kappa^*_V$ are rotativity and convertibility, respectively. 
These coefficients for a relativistic plasma with a power-law energy
distribution have been derived by Sazonov (1969a,b) for a
particular frequency region, $\nu_{min} \ll \nu$.
We have extended the frequency region to $\nu \ll \nu_{min}$,
and both cases are summarized in Appendix \ref{sec:app}.
For the emissivity $\eta_{(I,Q,U,V)}$, 
we consider only the synchrotron emission.

There are mainly two ways to generate the circularity $V$ from
synchrotron sources.
The first one is intrinsic emission due to the
emissivity $\eta_{V}$ (Legg \& Westfold 1968).
The second is FC (e.g., Jones \& O'Dell 1977a),
which is the conversion of $Q$ and $U$ into $V$ 
by means of $\kappa_{Q}^{*}$ in equation (\ref{eq:trans}).
If the natural modes of a plasma 
are nearly circular, the LP vector of propagating radiation rotates
since the left- and right-circular modes have different phase velocities
due to birefringence. This effect 
is well known as Faraday Rotation (FR) (Rybicki \& Lightman 1979).
FC is a similar phenomenon caused by birefringence of a medium.
If the natural modes are linearly or elliptically polarized,
the difference in the phase velocities leads to
the cyclic conversion between CP and LP.
FC becomes effective around the self-absorption
frequency since $|\kappa_{Q}^{*}/\kappa_{I}|\sim 1$ for $p\sim 2$
from equations (\ref{eq:coef2}) and (\ref{eq:coef3}).
Note that, if the magnetic field is uniform,
the synchrotron emission does not generate $U$
in the coordinate system with $\phi=0$.
Therefore FC does not occur without 
the rotation of $Q$ into $U$ by $\kappa_{V}^{*}$.

When we consider the tangled magnetic fields later,
the coefficients in equation (\ref{eq:trans}) are averaged with
respect to the distribution of the magnetic field.
This is justified when 
the typical scale over which the Stokes parameters change 
is much larger than that of the field orientation and hence
the correlations between the Stokes parameters
and the transfer coefficients tend to zero 
(Ruszkowski \& Begelman 2002).
\footnote{
In this Letter we also neglect the coupling of the natural modes 
in an inhomogeneous medium
(Jones \& O'Dell 1977b;
 for its validity see Ruszkowski \& Begelman 2002).}
\section{MAGNETIC FIELD CONFIGURATION}\label{sec:mag}
Since we know very little about the magnetic field configuration
in the afterglows of the GRB's,
we here consider two components, i.e.,
the ordered field and the axisymmetric tangled field.
There is a preferred direction in which the fluid moves.
Therefore, 
the assumption of the axisymmetry of the tangled field 
is quite natural.    
The ordered field is expected 
in the patchy model (Gruzinov \& Waxman 1999),
or if the magnetic field advected from the central engine prevails
or an ordered magnetic field exists in the medium into which the shock
propagates (Granot \& K$\ddot{\rm o}$nigl 2003).
The tangled field could be produced by the Weibel instability
(Medvedev \& Loeb 1999) or the turbulence behind the shock.

We set a coordinate system in the shocked fluid frame as shown in
Figure \ref{fig:magdir}.
The z-axis is the radial direction, in which the fluid moves.
The ordered field is characterized by the strength $B_{ord}$
and the direction $(\vartheta_{ord}, \varphi_{ord})$.
The axisymmetric tangled field is described by
the field strength as a function of $\vartheta_{tan}$, 
$B_{tan}(\vartheta_{tan})$,
and the probability per unit solid angle $f(\vartheta_{tan})$.
According to Sari (1999), we adopt
$B_{tan}(\vartheta_{tan})\propto (\xi^2 \sin^2\vartheta_{tan} 
+\cos^2 \vartheta_{tan})^{-1/2}$
and $f(\vartheta_{tan})\propto B_{tan}^3(\vartheta_{tan})$.
If $\xi \gg 1$, $\langle B_{\parallel}^{2} \rangle \gg
\langle B_{\perp}^{2} \rangle$, and vice versa,
where $B_{\parallel}$ and $B_{\perp}$ are the tangled magnetic field
components parallel and perpendicular to the z-axis, respectively.
We parametrize the ratio of the ordered field to the tangled one
by $\zeta=B_{ord}^{2}/(\langle B_{\parallel}^{2} \rangle +
\langle B_{\perp}^{2} \rangle)$.
The total strength $B^{2}=B_{ord}^{2}+\langle B_{\parallel}^{2} \rangle +
\langle B_{\perp}^{2} \rangle$ is determined by the afterglow model
in the next section.

Given a magnetic field geometry and the plasma parameters from 
the afterglow model, we can take an angular average of the coefficients
and solve the transfer equation (\ref{eq:trans}) to obtain
the Stokes parameters.
Although the shock structure may be important (e.g., Ioka 2003),
we neglect it as a first step.
We define the observed frequency
as the frequency in the fluid frame multiplied by 
the Lorentz factor of the fluid in the lab frame $\gamma$,
whereas ratios of the Stokes parameters are Lorentz invariant.

If $\zeta = 0$ (no ordered field),
we can immediately find that 
$\langle \eta_Q \sin 2 \phi \rangle$,
$\langle \eta_{V} \rangle$,
$\langle \kappa_Q \sin 2 \phi \rangle$,
$\langle \kappa^{*}_Q \sin 2 \phi \rangle$, 
$\langle \kappa_{V} \rangle$ and $\langle \kappa_{V}^{*} \rangle$
vanish as a result of the axisymmetry and reflection symmetry about
the $xy$ plane.
{\it Therefore the tangled field alone cannot generate CP}. 
Intuitively this is because there is always an oppositely directed pair
of magnetic fields, so that no handedness exists.

\section{APPLICATIONS TO GRB AFTERGLOWS}\label{sec:after}
First we consider the forward external shock with energy $E$
propagating into a constant surrounding density $n$.
According to the standard afterglow model (Sari, Piran \& Narayan 1998), 
the Lorentz factor of the shocked fluid
and the radius of the shock evolve as
$\gamma=(17 E/1024 \pi n m_{p} c^{5} t^{3})^{1/8}$ and 
$R=(17 E t/4\pi m_{p} n c)^{1/4}$, respectively,
where $t$ is the observer time.
The shell thickness in the shocked fluid frame can be estimated by $R/\gamma$,
which we use as the path length of the transfer equation 
(\ref{eq:trans}).
We assume that electrons are accelerated in the shock to 
a power law distribution of Lorentz factor $\gamma_e$,
$N(\gamma_e)d\gamma_e \propto \gamma_e^{-p} d\gamma_e$
for $\gamma_e>\gamma_{min}$,
where $N_{e}= \int_{\gamma_{min}}N(\gamma_e)d\gamma_e=4 \gamma n$
is the electron number density in the shocked fluid frame and $p>2$.
The minimum electron energy
and the magnetic field strength in the shocked fluid frame are given by
$\gamma_{min}=\epsilon_e\left[(p-2)/(p-1)\right](m_p/m_e) \gamma$
and $B=(32\pi m_p \epsilon_B n)^{1/2} \gamma c$, respectively,
where the parameters $\epsilon_e$ and $\epsilon_B$
are fractions of shock energy that go into the electrons
and the magnetic energy, respectively.
Thus we can calculate $\gamma$, $\gamma_{min}$, $B$, $R/\gamma$ 
and $N_{e}$
as a function of $t$ for given $E$, $n$, $p$, $\epsilon_{e}$
and $\epsilon_{B}$.
We adopt $E=10^{52}$ erg, $n=1$ proton cm$^{-3}$, $p=2.2$, $\epsilon_{e}=0.1$
and $\epsilon_{B}=0.01$ (Panaitescu \& Kumar 2002).
For simplicity, we temporarily neglect the jet effects.

The typical synchrotron frequency 
is $\nu_m=eB \gamma_{min}^2 \gamma/2\pi m_e c=145
\epsilon_{e,-1}^{2} \epsilon_{B,-2}^{1/2} E_{52}^{1/2} t_{1\rm day}^{-3/2}$ GHz
where the convention $Q=10^{x} Q_{x}$ is used except for 
$t_{1 {\rm day}}=t/1$ day.
For $\nu>\nu_{m}$ and $\nu<\nu_{m}$, we use equations (\ref{eq:coef2})
and (\ref{eq:coef3}), respectively.
By using equation (\ref{eq:coef3}),
we can estimate the self-absorption frequency from $\kappa_{I} R/\gamma=1$
as $\nu_{a}=41 \epsilon_{e,-1}^{-1} \epsilon_{B,-2}^{1/5} 
E_{52}^{1/5} n^{3/5}$ GHz,
where we put $\sin \theta=1$.
The flux is given by $F_{\nu} \propto t^{1/2} \nu^{\beta}$
with $\beta=2$ and $1/3$ for 
$\nu<\nu_{a}$ and $\nu_{a}<\nu<\nu_{m}$, respectively
(Sari, Piran \& Narayan 1998).
We can neglect the electron cooling for our interests.

\begin{figure}
\plotone{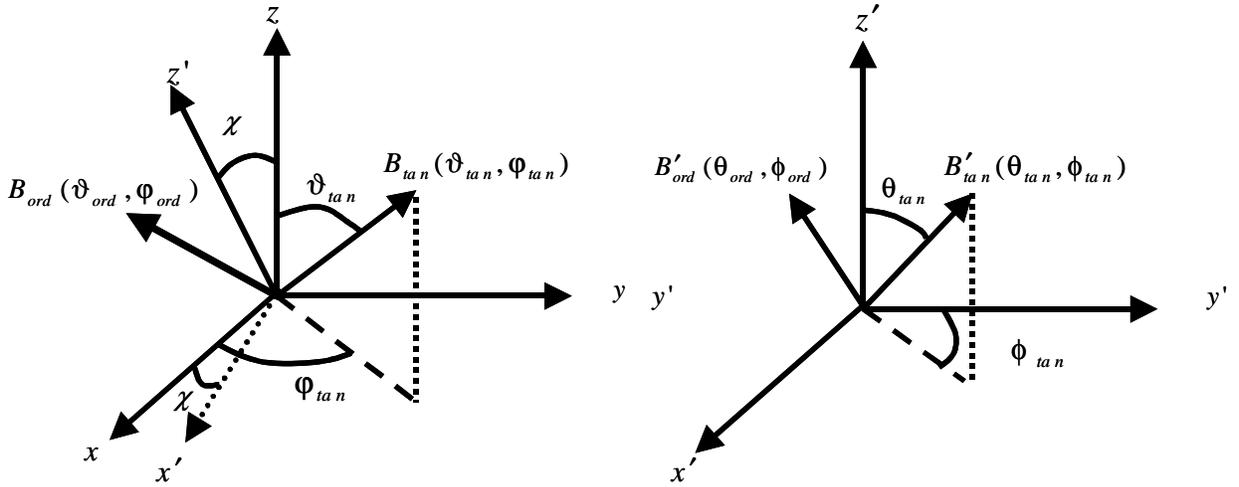}
\caption{
Magnetic field orientation in the shocked fluid frame. 
({\it left panel}) The fluid flow is along the $z$ axis. 
We introduce the ordered and tangled magnetic field. The former is
defined by $(\vartheta_{ord}, \varphi_{ord})$, 
while the latter is stochastic and
distributes as a function of ($\vartheta_{tan}$, $\varphi_{tan}$).
The observer is in the direction of the axis $z'$, which is
specified by the angle $\chi$ on the $xz$ plane. 
({\it right panel}) It is convienient to take a new coordinate in order
to deal with the radiation transfer. 
In this coordinate, the magnetic field direction is specified by 
($\theta$, $\phi$).}
\label{fig:magdir}
\end{figure}

\begin{figure}
\plotone{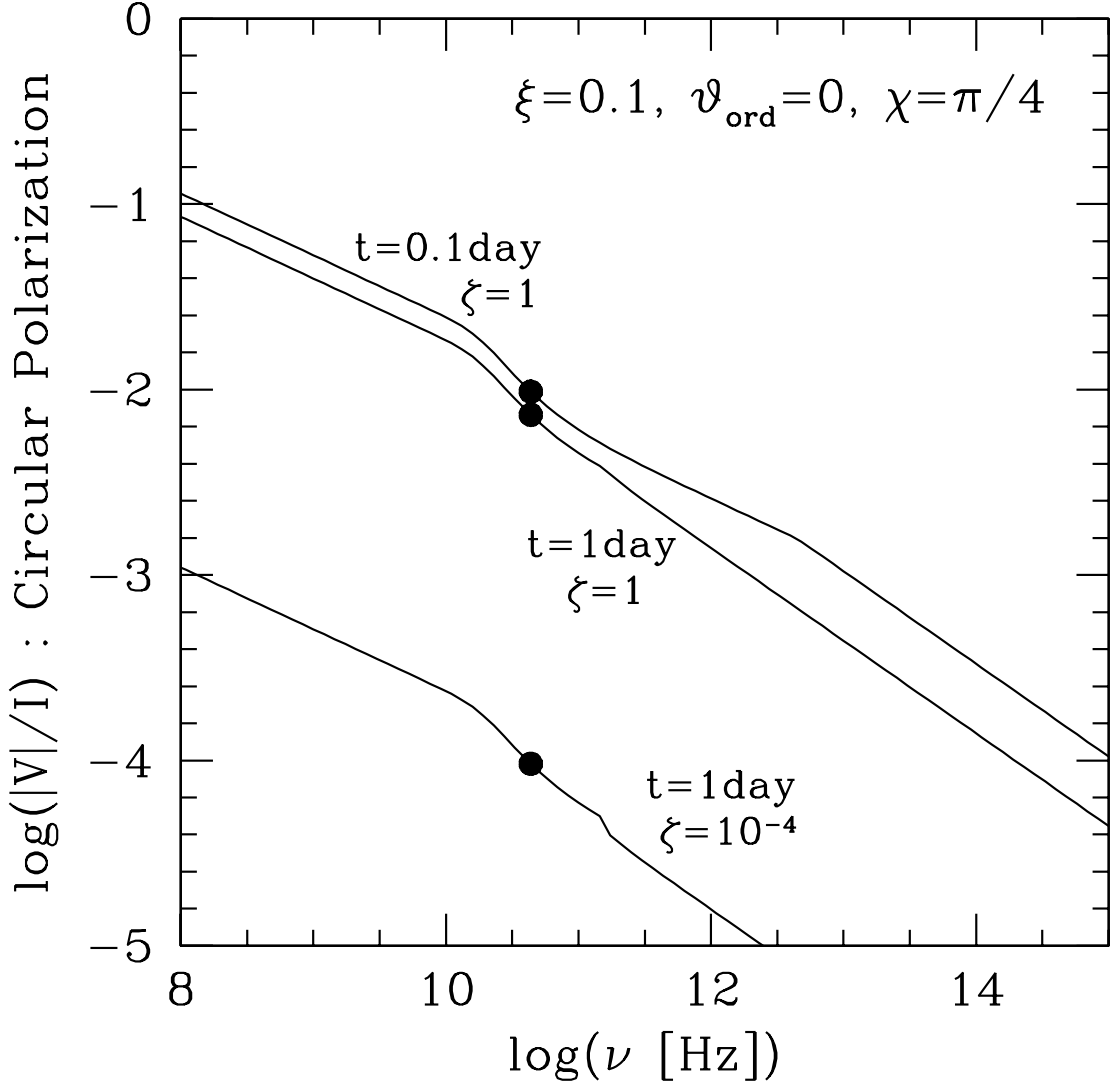}
\caption{
Circular polarization ($=|V|/I$) 
from a forward shock is shown as a function of the observed frequency $\nu$
for $t=1$ and $0.1$ day.
The self-absorption frequencies $\nu_{a}$ are marked by black points.
The ratio of the ordered field to the tangled one is defined by
$\zeta=B_{ord}^{2}/(\langle B_{\parallel}^{2} \rangle +
\langle B_{\perp}^{2} \rangle)$.
The ordered field is directed to $\vartheta_{ord}=0$.
The perpendicular tangled component dominates the parallel one, i.e., 
$\xi=0.1$.
The observer is in the direction $\chi=\pi/4$ (see Figure \ref{fig:magdir}).
For this configuration, $V<0$.
We adopt $E=10^{52}$ erg, $n=1$ proton cm$^{-3}$, 
$p=2.2$, $\epsilon_{e}=0.1$ and $\epsilon_{B}=0.01$.
}
\label{fig:cpfor}
\end{figure}

Figure \ref{fig:cpfor} illustrates CP from a forward shock.
The degree of CP is about 1\% at the self-absorption frequency $\nu_{a}$
and $0.01\%$ at optical for $\zeta \simg 1$.
We have found that the dependence of CP on $t$ and $\xi$ is weak,
and the dependence on $\zeta$, $\vartheta_{ord}$, $\varphi_{ord}$
and $\chi$ is roughly $\propto \sqrt{\zeta/(1+\zeta)} \cos \theta_{ord}$
(see Figure \ref{fig:magdir} for the relation between 
$\theta_{ord}$, $\vartheta_{ord}$, $\varphi_{ord}$ and $\chi$).
The CP is mainly due to intrinsic emission.

Next we consider the reverse shock propagating the ejected shell itself
(M${\acute {\rm e}}$sz${\acute {\rm a}}$ros \& Rees 1997;
Sari \& Piran 1999a).
Let $\gamma_{0}$ and $T$ be an initial Lorentz factor of the shell
and the burst duration, respectively.
Then the Lorentz factor at the shock crossing time is given by
$\gamma_{\times}=\min (\gamma_{0}, \gamma_{c})$,
where $\gamma_{c} \equiv (3 E/32\pi n m_{p} c^{5} T^{3})^{1/8}$
is a critical Lorentz factor
(Kobayashi 2000; Kobayashi \& Zhang 2003; 
Zhang, Kobayashi \& M${\acute {\rm e}}$sz${\acute {\rm a}}$ros 2003;
Sari \& Piran 1999a,b). The shock crossing time is given by 
$t_{\times}=(\gamma_{0}/\gamma_{\times})^{8/3} T$.
At $t=t_{\times}$ the minimum electron energy,
the magnetic field strength and the electron number density
in the shocked fluid frame are given by
$\gamma_{min}=\epsilon_e\left[(p-2)/(p-1)\right](m_p/m_e) 
\gamma_{0}/\gamma_{\times}$,
$B=(32\pi m_p \epsilon_B n)^{1/2} \gamma_{\times} c$
and $N_{e}=4\gamma_{\times}^{3} n/\gamma_{0}$, respectively.
After the shock crossing, these quantities approximately evolve as
$\gamma \propto t^{-7/16}$, $\gamma_{min}\propto t^{-13/48}$,
$B\propto t^{-13/24}$ and $N_{e} \propto t^{-13/16}$,
respectively (Kobayashi 2000; Kobayashi \& Sari 2000).
The shell thickness in the shocked fluid frame is $R/\gamma$
where $R=(17E t/4\pi m_{p} n c)^{1/4}$.
We adopt $\gamma_{0}=100$ and $T=100$ s.

The typical synchrotron frequency is 
$\nu_{m}=0.035$ 
$\epsilon_{e,-1}^{2}$ 
$\epsilon_{B,-2}^{1/2} n^{1/2}$
$t_{1day}^{-73/48}$ 
$T_{2}^{73/48}$ 
$\gamma_{0,2}^{2}$
$\max [1,(\gamma_{0}/\gamma_{c})^{73/18}]$ GHz.
Using equation (\ref{eq:coef2})
we can estimate the self-absorption frequency from $\kappa_{I} R/\gamma=1$ as
$\nu_{a}=20 \epsilon_{e,-1}^{2(p-1)/(p+4)}$
$\epsilon_{B,-2}^{(p+2)/2(p+4)}$
$n^{(p+5)/2(p+4)}$
$t_{1day}^{-(73p+122)/48(p+4)}$
$T_{2}^{(73p+146)/48(p+4)}$
$\gamma_{0,2}^{2}$
$E_{52}^{1/2(p+4)}$
$\max [1,(\gamma_{0}/\gamma_{c})^{(73p-70)/18(p+4)}]$
GHz, where we put $\sin \theta = 1$.
The flux peaks at $\nu \sim \nu_{a}$.
We can neglect the electron cooling for our interests.

\begin{figure}
\plotone{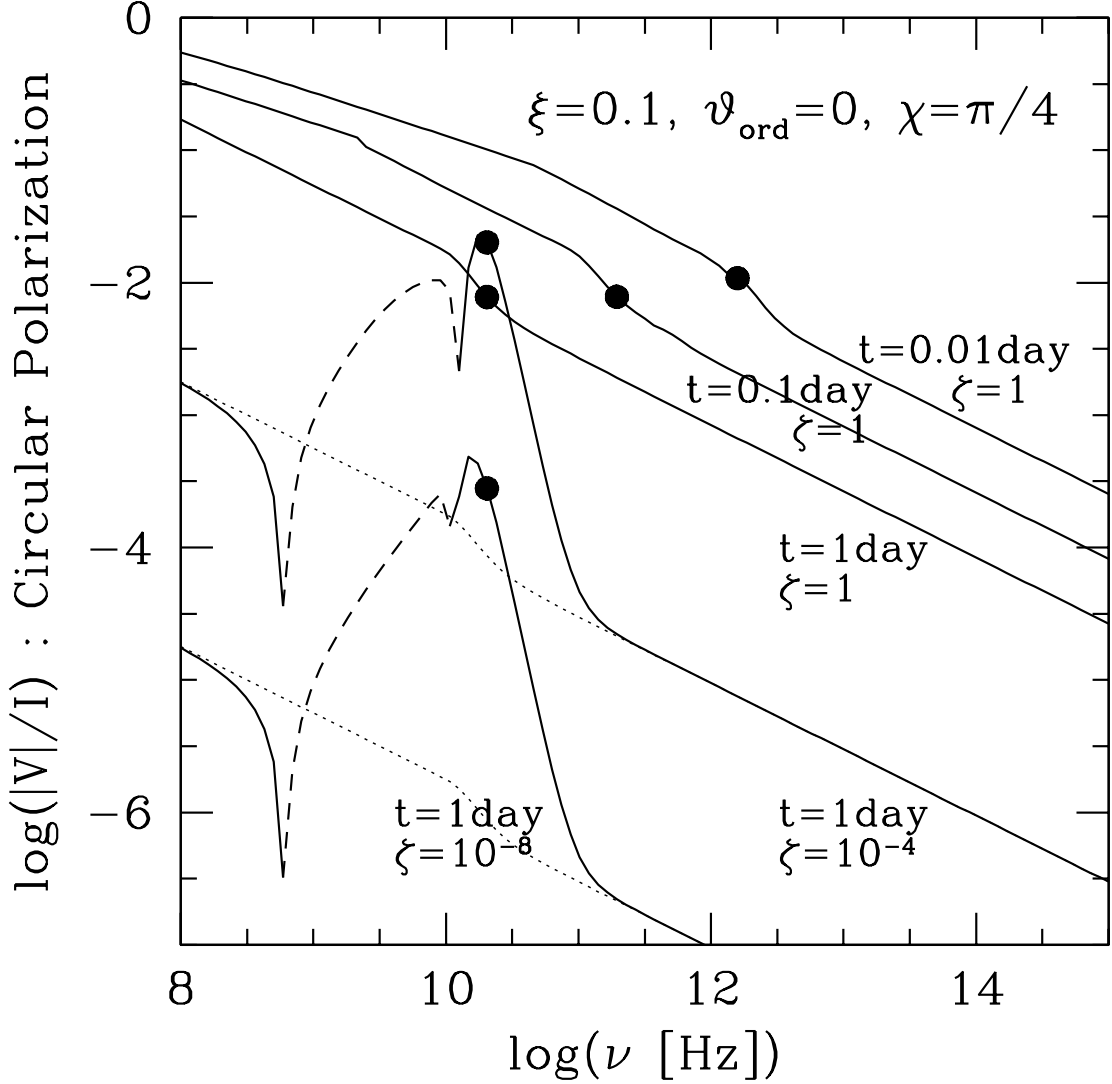}
\caption{
Circular polarization ($=|V|/I$) 
from a reverse shock is shown as a function of the observed frequency $\nu$
for $t=1$, $0.1$ and $0.01$ day.
For $V<0$ ($V>0$) we use solid (dashed) lines.
The self-absorption frequencies $\nu_{a}$ are marked by black points.
The dotted lines are calculated by putting $\kappa_{Q}^{*}=0$
(no conversion).
The ratio of the ordered field to the tangled one is defined by
$\zeta=B_{ord}^{2}/(\langle B_{\parallel}^{2} \rangle +
\langle B_{\perp}^{2} \rangle)$.
We adopt $E=10^{52}$ erg, $n=1$ proton cm$^{-3}$, 
$p=2.2$, $\epsilon_{e}=0.1$, $\epsilon_{B}=0.01$,
$\gamma_{0}=100$, $T=100$ s, $\vartheta_{ord}=0$, $\xi=0.1$ and $\chi=\pi/4$.
}
\label{fig:cprev}
\end{figure}

Figure \ref{fig:cprev} shows CP from a reverse shock.
For $\zeta \simg 1$, the degree of CP 
reaches $10$-$1$\% at radio frequencies,
and $0.1$-$0.01$\% at optical.
Even if $\zeta \sim 10^{-4}$, CP remains to be $\sim$ 1\%
at the self-absorption frequency $\nu_{a}$.
This is because of the FC.
The intrinsic CP decreases as $\zeta$ decreases,
but $V$ is generated from $Q$ and $U$ due to the FC.
The dependence of CP on $\xi$ is weak.
If FC is not effective,
the dependence on $\zeta$, $\vartheta_{ord}$, $\varphi_{ord}$
and $\chi$ is roughly $\propto \sqrt{\zeta/(1+\zeta)} \cos \theta_{ord}$,
while if FC is effective, it is not so simple.

So far we have calculated CP at one point on the afterglow image
on the sky.
The integration over the entire emitting region may suppress
the observed CP as in the case of LP (Sari 1999; Ghisellini \& Lazzati 1999).
Let us estimate the suppression factor when we observe 
a jet with an opening angle $\Theta_{0}$ from a viewing angle $\Theta_{v}$.
We consider the ordered magnetic field defined by 
$(\vartheta_{ord}, \varphi_{ord})$
in the coordinate with $x$-axis being the direction from the jet center
to the line of sight on the sky
and with $z$-axis being the direction in which the ejecta moves
(see Figure \ref{fig:magdir}).
We assume that the viewable region is a uniform disk 
of angular extent $1/\gamma$
centered around the line of sight to the observer,
since the afterglow image is rather homogeneous
for $\nu \siml \nu_{m}$ (Granot, Piran \& Sari 1999a,b)
or after the jet break (Ioka \& Nakamura 2001).
Then, if we assume $V/I \propto \cos \theta_{ord}$ and take the
coordinate origin at the line of sight on the sky,
the suppression factor is given by
$\int d\Theta d\Phi
\left[\cos\vartheta_{ord} \cos \chi_{\Theta} + 
\sin \chi_{\Theta} \sin \vartheta_{ord}
\cos (\varphi_{ord}-\Phi)\right]/ \int d\Theta d\Phi$,
where $\sin \chi_{\Theta}=2 \gamma \Theta/(1+\gamma^2 \Theta^2)$
and $\Theta_{v}^2+2\Theta_{v} \Theta \cos\Phi +\Theta^2<\Theta_{0}^{2}$.
Interestingly, no cancellation takes place 
when $\vartheta_{ord}=0$ or $\pi$.
Even when the cancellation occurs, e.g, 
in the case of $\vartheta_{ord}=\pi/2$,
some amount of CP remains if the visible region has an asymmetry 
due to the jet geometry
as in the case of LP.

\acknowledgments
We are grateful F.~Takahara for useful comments.
This work was supported in part by
Grant-in-Aid for Scientific Research 
of the Japanese Ministry of Education, Culture, Sports, Science
and Technology, No.00660 (KI).

\appendix
\section{Transfer coefficients}\label{sec:app}
We summarize the transfer coefficients in equation (\ref{eq:trans})
at an angle $\theta$ to the magnetic field $B$
(Sazonov 1969a,b; Sazonov \& Tsytovich 1968; Melrose 1980).
In this section we measure all quantities in the shocked fluid frame.
We assume that the electron number density in the interval 
of the Lorentz factor $d\gamma_{e}$ is power-law
$dN_{e}=N(\gamma_{e}) d\gamma_{e}
= \tilde{N}_e \gamma_{e}^{-p} d\gamma_{e}$
for $\gamma_{min} \leq \gamma_{e}$ with $p>2$.
Then, at frequencies $\nu_{min}\equiv \gamma^2_{min} \nu_{\perp} \ll \nu$
the coefficients are given by
\begin{eqnarray}
\eta_I &=& \eta_{\alpha}\eta_{\perp}(\nu/\nu_{\perp})^{-\alpha}, \quad
\eta_Q = \eta^Q_{\alpha}\eta_{\perp}(\nu/\nu_{\perp})^{-\alpha}, \quad
\eta_V = -\eta^V_{\alpha}\eta_{\perp}(\nu/\nu_{\perp})^{-\alpha-1/2}
\cot\theta, \nonumber\\
\kappa_I &=& \kappa_{\alpha}\kappa_{\perp}(\nu/\nu_{\perp})^{-\alpha-5/2}, 
\quad
\kappa_Q = \kappa^Q_{\alpha}\kappa_{\perp}(\nu/\nu_{\perp})^{-\alpha-5/2}, 
\quad
\kappa_V = \kappa^V_{\alpha}\kappa_{\perp}(\nu/\nu_{\perp})^{-\alpha-3} 
\cot \theta, \nonumber\\
\kappa^*_Q &=& -\kappa^{*Q}_{\alpha}\kappa_{\perp}(\nu/\nu_{\perp})^{-3} 
\gamma^{-2\alpha+1}_{min}\{1-(\nu/\nu_{min})^{-\alpha+1/2}\}, 
\nonumber\\  
\kappa^*_V &=& \kappa^{*V}_{\alpha}\kappa_{\perp}
(\nu/\nu_{\perp})^{-2}(\ln \gamma_{min})\gamma^{-2(\alpha+1)}_{min}
\cot \theta,
\label{eq:coef2}
\end{eqnarray}
where $\nu_{\perp} \equiv |e| B \sin \theta/2\pi m_e c$,
$\eta_{\perp} \equiv (e^2/c)\tilde{N}_e \nu_{\perp}$,
$\kappa_{\perp} \equiv (e^2/m_{e} c) \tilde{N}_e/\nu_{\perp}$,
$\alpha \equiv (p-1)/2$, and
\begin{eqnarray}
\eta_{\alpha} &=& 
\frac{3^{\alpha+1/2}}{4(\alpha+1)}
~\Gamma \left(\frac{\alpha}{2}+\frac{11}{6} \right)
\Gamma \left(\frac{\alpha}{2}+\frac{1}{6} \right),\quad
\eta^Q_{\alpha} =
\frac{3^{\alpha+1/2}}{4(\alpha+5/3)}
~\Gamma \left(\frac{\alpha}{2}+\frac{11}{6} \right)
\Gamma \left(\frac{\alpha}{2}+\frac{1}{6} \right),\nonumber\\
\eta^V_{\alpha} &=& 
\frac{3^{\alpha}(\alpha+3/2)}{2(\alpha+1/2)}
~\Gamma \left(\frac{\alpha}{2}+\frac{11}{12} \right)
\Gamma \left(\frac{\alpha}{2}+\frac{7}{12} \right),\quad
\kappa_{\alpha} = 
\frac{3^{\alpha+1}}{4}
~\Gamma \left(\frac{\alpha}{2}+\frac{25}{12} \right)
\Gamma \left(\frac{\alpha}{2}+\frac{5}{12} \right),\nonumber\\
\kappa^{Q}_{\alpha} &=& 
\frac{3^{\alpha+1}}{4}
\frac{\alpha+3/2}{\alpha+13/6}
~\Gamma \left(\frac{\alpha}{2}+\frac{25}{12} \right)
\Gamma \left(\frac{\alpha}{2}+\frac{5}{12} \right),\nonumber\\
\kappa^{V}_{\alpha} &=& 
\frac{3^{\alpha+1/2}}{2}
\frac{\alpha+2}{\alpha+1}\left(\alpha+\frac{3}{2}\right)
~\Gamma \left(\frac{\alpha}{2}+\frac{7}{6} \right)
\Gamma \left(\frac{\alpha}{2}+\frac{5}{6} \right),\nonumber\\
\kappa^{* Q}_{\alpha} &=& \frac{\alpha+3/2}{4(\alpha-1/2)},\quad
\kappa^{* V}_{\alpha} =
2~\frac{\alpha+3/2}{\alpha+1}.
\end{eqnarray}
The above notations are the same as Jones \& O'Dell
(1977a; But $\kappa^{* Q}_{\alpha}$ is different.
The appropriate integration of equation (9) in Sazonov (1969a) 
gives the above value).
For the frequency region, $\nu \ll \nu_{min}$, 
the analogous representations have not been derived yet. 
Such a frequency region becomes important in the application 
to the forward shock of GRB afterglow. 
In this case, we obtain the following expressions:
\begin{eqnarray}
\eta_I &=& \eta_{\alpha}\eta_{\perp}(\nu/\nu_{\perp})^{1/3} 
\gamma^{-2\alpha-2/3}_{min},\quad
\eta_Q = \frac{1}{2} \eta_I, \quad
\eta_V = -\eta^{V}_{\alpha}\eta_{\perp}\gamma^{-2\alpha-1}_{min}
\cot \theta,\nonumber\\
\kappa_I &=& \kappa_{\alpha}\kappa_{\perp}(\nu/\nu_{\perp})^{-5/3} 
\gamma^{-2\alpha-5/3}_{min},\quad
\kappa_Q = \frac{1}{2} \kappa_I, \quad
\kappa_V = \kappa^{V}_{\alpha}\kappa_{\perp}(\nu/\nu_{\perp})^{-2}
\gamma^{-2(\alpha+1)}_{min}
\cot \theta,\nonumber\\
\kappa^{*}_{Q} &=& \kappa^{* Q}_{\alpha}\kappa_{\perp}(\nu/\nu_{\perp})
^{-5/3}\gamma^{-2\alpha-5/3}_{min}, \quad
\kappa^{*}_{V} = \kappa^{* V}_{\alpha}\kappa_{\perp}(\nu/\nu_{\perp})
^{-2}(\ln \gamma_{min})\gamma^{-2(\alpha+1)}_{min} \cot \theta, 
\label{eq:coef3}
\end{eqnarray}
where
\begin{eqnarray}
\eta_{\alpha} &=& 
\frac{3^{1/6}}{2(\alpha+1/3)}
~\Gamma \left(\frac{2}{3} \right),\quad
\eta^V_{\alpha} = 
\frac{\pi}{3(\alpha+1/2)},\quad
\kappa_{\alpha} = 
\frac{3^{1/6}}{2}\frac{\alpha+3/2}{\alpha+5/6}
~\Gamma \left(\frac{2}{3} \right),\nonumber\\
\kappa^{V}_{\alpha} &=& \frac{\pi(\alpha+3/2)}{3(\alpha+1)},\quad
\kappa^{* Q}_{\alpha} =
\frac{3^{-1/3}}{4}\frac{\alpha+3/2}{\alpha+5/6}
~\Gamma \left(\frac{2}{3} \right),\quad
\kappa^{* V}_{\alpha} =
2~\frac{\alpha+3/2}{\alpha+1}.
\end{eqnarray}   
Note that the intrinsic LP ($\eta_{Q}/\eta_{I}$) in this frequency region 
is not $(p+1)/(p+7/3) \approx 70\%$ but $50\%$.
In GRB 021206, considerable photons are below the break energy, i.e., 
$\nu \ll \nu_{min} $ (Coburn \& Boggs 2003), so that even an ordered
field might not explain the observed LP $\approx 80\pm 20 \% $.


%
%

\newpage

\newpage

\newpage

\end{document}